\documentclass[aps,prd,tightenlines,showpacs,amsmath,amssymb]{revtex4}%
\usepackage{graphicx}
\usepackage{epsfig}
\usepackage{graphics,color}
\usepackage{amssymb}
\usepackage{amsmath}
\usepackage{latexsym}
\usepackage{hyperref}
\usepackage{amsbsy}
\usepackage{bm}
\usepackage{url}

\begin{document}
\title{Pion String evolving in a thermal bath}
\author{Fan Lu}
\author{Qichang Chen}
\author{Hong Mao}
\email {mao@hznu.edu.cn (corresponding author)}
\address{Department of Physics, Hangzhou Normal University, Hangzhou 310036, China}


\begin{abstract}
By using the symmetry improved CJT effective formalism, we study a pion string of the $O(4)$ linear sigma model at finite temperature in chiral limit. In terms of the Kibble-Zurek mechanism we reconsider the production and evolution of the pion string in a thermal bath. Finally, we estimate the pion string density and its possible signal during the chiral phase transition.
\end{abstract}


\maketitle

\section{Introduction}
Topological defects produced at phase transition are known to play an important role in early universe cosmology\cite{Villenkin00}\cite{Hindmarsh:1994re}\cite{Manton:2004tk}. On the one hand, the early evolution of the universe undergoes a sequence of phase transitions and topological defects in these phase transitions may have observable consequences to the properties of the universe today. Such as the cosmic strings, they have been suggested as one possible source for the primordial density perturbations that give rise to the large-scale structure of the universe and the temperature fluctuations of the cosmic microwave background (CMB) radiation\cite{Villenkin00},  they could also contribute to structure formation or generate primordial magnetic fields which are coherent on cosmological scales\cite{Brandenberger:1998ew}. On the other hand, some predictive topological defects in the particle physics models, such as domain walls, which have problematic and cause a conflict with cosmology, can be ruled out\cite{Eto:2013bxa}.

The linear sigma model \cite{GellMann:1960np} for the phenomenology of QCD has been proposed to describe the vacuum structure with incorporating chiral symmetry and its spontaneous breaking. The model can be used to describe a chiral phase transition in quantum chromodynamics (QCD) at finite temperature \cite{Rischke:2003mt,Roh:1996ek,AmelinoCamelia:1997dd,Petropoulos:1998gt,Lenaghan:1999si,Lenaghan:2000ey,Roder:2003uz,Phat:2003eh,Nemoto:1999qf,Petropoulos:2004bt,Mao:2006zr} within the Cornwall-Jackiw-Tomboulis (CJT) formalism\cite{Cornwall:1974vz}. Similarly to the standard model of particle physics, traditionally, the linear sigma model does not give rise to topological defects which are stable in vacuum. However, if certain of the fields were constrained to vanish,  it is still possible to construct a string-like configuration in the linear sigma model, the pion string, which would be topological defects\cite{Zhang:1997is}. This kind of string is usually named as ``embedded defects"\cite{Achucarro:1999it} , which is not topological stable because any field configuration can be continuously deformed to the trivial vacuum. In order to cure this problem, a plasma stabilization mechanism had been proposed in Ref.\cite{Nagasawa:1999iv}\cite{Nagasawa:2002at}. In their works, they argued that the interaction of the pion fields with the charged plasma generates a correction to the effective potential and this correction reduces the vacuum manifold $S^3$ of the zero temperature theory to a lower dimensional submanifold $S^1$, which makes the pion string stable. Moreover, this suggested stabilization mechanism for the pion string has been put on a firmer confirmation in recent works by Karouby and Brandenberger\cite{Karouby:2012yz}\cite{Karouby:2013vza}. So that we have confidence to believe that pion strings are expected to be formed during the QCD phase transition in the early universe as well as in experiment of the heavy ion collisions.

In order to study whether the pion string exists during the chiral phase transition, e.g. in the heavy ion collisions or in early universe. Firstly, we need to know how to estimate the production rate of pion strings in chiral phase transition, thereafter, we should know how to study the evolution and decay of these pion strings in a thermal bath when cooling from high temperature. Finally, we can reveal their possible trails in the experiment observations. So far the theoretical scenario that can be applied to study the formation of topological defects in rapid phase transition with global symmetry is the Kibble-Zurek mechanism\cite{Kibble76,Zurek:1996sj,Mao:2004ym,delCampo:2013nla}. However, the CJT effective action in chiral limit presented in Refs.\cite{Rischke:2003mt,Roh:1996ek,AmelinoCamelia:1997dd,Petropoulos:1998gt,Lenaghan:1999si,Lenaghan:2000ey,Roder:2003uz,Phat:2003eh,Nemoto:1999qf,Petropoulos:2004bt,Mao:2006zr} violates the Goldstone theorem and gives massive pions in the spontaneous symmetry breaking phase, besides, the temperature-dependent order parameter predicts a first order phase transition. All these not only disagree with the rigorous universality arguments, where the chiral phase transition can be of second order if the $U(1)_A$ symmetry is explicitly broken by instanton for $N_f=2$ flavors of massless quarks \cite{Pisarski:1983ms}, but also conflict  with recent results reported by Chiku and Hatsuda\cite{Chiku:1997va}\cite{Chiku:1998kd} where by employing the optimal perturbation theory (OPT) and the real time formalism, they proved that the Goldstone theorem is always satisfied in any given order of the loop expansion in OPT in the $O(4)$ linear sigma model. Therefore, the direct use of the Kibble-Zurek mechanism to the model has broken down.

In order to restore the Goldstone theorem in chiral limit, Pilaftsis and Teresi recently developed a novel symmetry improved CJT formalism by consistently encoding global symmetries in loop-wise expansions or truncations of the CJT effective action\cite{Pilaftsis:2013xna}. Furthermore, this formalism has proven to avoid the existing problems and keep the features of the second order phase transition in the $O(4)$ linear sigma model\cite{Mao:2013gva}. Thus, in this article we will apply the Kibble-Zurek mechanism to study the formation and evolution of the pion string embedded in a quark-gluon plasma at the LHC Pb-Pb collisions with a beam energy at $\sqrt{s_{NN}}=2760 \mathrm{GeV}$ by adopting the symmetry-improved CJT effective potential.

The organization of this paper is as follows. In the next section we give a brief review of the pion string in the linear sigma model. In Sect. III, we calculate the effective potential by using the symmetry improved CJT formalism and obtain new equation of motion for the pion string. In Sect. IV, by using the Kibble-Zurek mechanism, the production and evolution of the pion string at finite temperature have been discussed. In Sect.V we discuss decays of pion strings for both the stable and unstable cases and we also address their possible observational consequence in experiment. Finally, Section VI summarizes our conclusions and discusses some open problems.

\section{The pion string in the linear sigma model}
The Lagrangian density of the $SU(2)_{R}\times SU(2)_{L}$ symmetry linear sigma model has the form
\begin{equation}\label{Eq:Lag1}
{\cal L}=\frac{1}{2} \left(\partial _{\mu}\sigma \partial ^{\mu}\sigma +
\partial _{\mu}\vec{\pi} \cdot \partial ^{\mu}\vec{\pi}\right)
-U(\sigma ,\vec{\pi}) ,
\end{equation}
where the potential for the $\sigma$ and $\vec{\pi}$ is parameterized as
\begin{equation}\label{Eq:Potential1}
U(\sigma,
\vec{\pi})=\frac{m^2}{2}(\sigma^2+\vec{\pi}^2)+\frac{\lambda}{24}(\sigma^2+\vec{\pi}^2)^2.
\end{equation}

In our discussion of the pion string it proves convenient to define the new fields
\begin{eqnarray}
\varphi =\frac{\sigma +i\pi ^0}{\sqrt{2}}
\end{eqnarray}
and
\begin{eqnarray}
 \pi ^{\pm }=\frac{\pi^1\pm i\pi ^2}{ \sqrt{2}}.
\end{eqnarray}
Then, the Lagrangian in Eq.(\ref{Eq:Lag1}) in terms of the $\varphi$ and $\pi ^{\pm }$ fields is of the form
\begin{equation}\label{eq2}
{\mathcal{L}}_\Phi =(\partial _\mu \varphi ^{*})(\partial ^\mu \varphi
)+(\partial _\mu \pi
^{+})(\partial ^\mu \pi ^{-})-\chi (\varphi ^{*}\varphi +\pi ^{+}\pi ^{-}-\frac{%
f_\pi ^2}2)^2,
\end{equation}
with $\chi=\lambda/6$.
The time-independent equations of motion read:
\begin{equation}\label{pionsf}
\nabla ^2\varphi =2\chi (\varphi ^{*}\varphi +\pi ^{+}\pi ^{-}-\frac{f_\pi ^2}2%
)\varphi
\end{equation}
and
\begin{equation}
\nabla ^2\pi^+ =2\chi (\varphi ^{*}\varphi +\pi ^{+}\pi
^{-}-\frac{f_\pi ^2}2)\pi^+.
\end{equation}
The global pion string extending linearly to the $z$-axis is obtained by solving the equations of motion with an antisymmetric vortex ansatz in the cylindrical coordinates in $x$-$y$ plane, given by\cite{Zhang:1997is}
\begin{subequations}\label{pss}
\begin{eqnarray}
 \varphi &=& \frac{f_{\pi}}{\sqrt{2}}\rho(r)e^{in\theta}, \\
 \pi ^{\pm }&=& 0.
\end{eqnarray}
\end{subequations}
Here, the coordinates $r$ and $\theta$ are polar coordinates in
$x-y$ plane and the integer $n$ is the winding number. In the
following discussion, we will restrict ourselves to $n=1$.
Plugging the ansatz (\ref{pss}) into the above equations of motion, we get a second order differential equation
\begin{eqnarray}
\rho''+\frac{\rho'}{r}-\frac{\rho}{r^2}=\chi f^2_{\pi}(\rho^2-1)\rho,
\end{eqnarray}
with the boundary conditions
\begin{equation}
\lim_{r\rightarrow \infty}\rho(r)=1, \qquad  \lim_{r\rightarrow 0}\rho(r)=0.
\end{equation}

The energy per unit length of the string can be expressed as
\begin{eqnarray}\label{energy}
E=\pi f^2_{\pi}\int^{\infty}_0 dr r \left[ \rho'^2+\frac{\rho^2}{r^2}+\frac{\chi}{2}(\rho^2-1)^2 \right],
\end{eqnarray}
where the first two terms come from the derivative of the field $\varphi$ and the last term is from the potential of the $\sigma$ and $\vec{\pi}$ fields. Since the kinetic energy is logarithmically divergent, we need to introduce a cutoff $R$ in Eq.(\ref{energy}) which is the horizon size or the typical separation length between strings. Such a typical distance between strings should  be determined by the initial string number density at the formation during the chiral phase transition based on the Kibble-Zurek mechanism. In general, $R=O(\mathrm{fm})$.

At tree level and zero temperature the parameters of the Lagrangian are fixed in a way that these masses agree with the observed value of pion masses and the most commonly accepted value for sigma mass. In the following numerical calculation, we take $m_{\sigma}=500$ MeV and $f_{\pi}=93$ MeV as typical values. In the chiral limit, the coupling constant $\lambda$ is chosen to be $\lambda=3 m^2_{\sigma}/f^2_{\pi}$.

\section {Effective potential computation}
A convenient framework of studying chiral phase transition is the quantum thermal field theory. Within this framework, a resummation method introduced by Cornwall, Jackiw and Tomboulis (CJT) is an important and useful theoretical tool\cite{Cornwall:1974vz}. However, in most of studies relevant to the truncated CJT formalism, there are two major drawbacks. One is that, in chiral limit, the CJT effective action violates the Goldstone theorem and gives massive pions in the spontaneous symmetry breaking phase. The other is that the temperature-dependent order parameter predicts a first order phase transition. This prediction, of course, disagrees with the rigorous universality arguments, where the chiral phase transition can be of second order if the $U(1)_A$ symmetry is explicitly broken by instanton for $N_f=2$ flavors of massless quarks\cite{Pisarski:1983ms}. While the former conclusion comes into conflict with recent results reported by Chiku and Hatsuda\cite{Chiku:1997va}\cite{Chiku:1998kd}, where by employing the optimal perturbation theory (OPT) and the real time formalism, they proved that the Goldstone theorem is always satisfied in any given order of the loop expansion in OPT in the $O(4)$ linear sigma model. Therefore, in following discussions, we take the symmetry-improved CJT effective as a prototype framework.

In the case of the linear sigma model,  the CJT effective potential at finite temperature will appear as\cite{Cornwall:1974vz}\cite{Pilaftsis:2013xna} \cite{Mao:2013gva}
\begin{eqnarray}\label{epotential0}
V(\phi,G_{\sigma}, G_{\pi}) &=& U(\phi)+\frac{1}{2}\int_{\beta}\ln
G^{-1}_{\sigma}(\phi;k)+\frac{3}{2}\int_{\beta}\ln
G^{-1}_{\pi}(\phi;k)
\nonumber\\&&+\frac{1}{2}\int_{\beta}\left[D^{-1}_{\sigma}(\phi;k)G_{\sigma}(\phi;k)-1 \right]
+\frac{3}{2}\int_{\beta}\left[ D^{-1}_{\pi}(\phi;k)G_{\pi}(\phi;k)-1 \right]
\nonumber\\&& +V_2(\phi,
G_{\sigma}, G_{\pi}).
\end{eqnarray}
Here, after shifting the sigma field as $\sigma\rightarrow \sigma+\phi$, the classical potential  $U(\phi)$ has the form
\begin{equation}
U(\phi)=\frac{1}{2}m^2\phi^2+\frac{\lambda}{24}\phi^4,
\end{equation}
and the last term $V_2(\phi, G_{\sigma}, G_{\pi})$ denotes the contribution from the ``$\infty$" diagrams, which is equivalent to the Hartree approximation, explicitly,
\begin{eqnarray}
V_2(\phi, G_{\sigma}, G_{\pi}) &=& 3\frac{\lambda}{24}
\left[\int_{\beta}G_{\sigma}(\phi;k)
\right]^2+6\frac{\lambda}{24}\left[\int_{\beta}G_{\sigma}(\phi;k)\right]\left[\int_{\beta}G_{\pi}(\phi;k)
\right] \nonumber \\&&
+15\frac{\lambda}{24} \left[\int_{\beta}G_{\pi}(\phi;k)\right]^2.
\end{eqnarray}
Here and there, we use the imaginary-time formalism to compute quantities at nonzero temperature, our notation is
\begin{eqnarray}
\int\frac{d^4k}{(2\pi)^4}f(k) \rightarrow \frac{1}{\beta}\sum_n
\int\frac{d^3\vec{k}}{(2\pi)^3}f(i\omega_n,\vec{k}) \nonumber
\equiv \int_{\beta}f(i\omega_n,\vec{k}),
\end{eqnarray}
where $\beta$ is the inverse temperature, $\beta=\frac{1}{k_BT}$, and as usual Boltzmann's constant is taken as $k_B=1$, and
$\omega_n=2\pi nT$, $n=0, \pm1, \pm2, \pm3, \cdots$. For simplicity we have introduced a subscript $\beta$ to denote integration and summation over the Matsubara frequency sums. Also, we denote $D_{\sigma}$ and $D_{\pi}$ as the tree-level sigma and pion propagators, while $G_{\sigma}$ and $D_{\pi}$ are treated as the full propagators. By minimizing the effective potential with respect to full propagators, we obtain a set of nonlinear gap equations for the sigma and pion thermal effective masses
\begin{subequations}\label{e-mass}
\begin{eqnarray}
M_{\sigma}^2 &=&
m^2+\frac{\lambda}{2}\phi^2+\frac{\lambda}{2}F(M_{\sigma})
+\frac{\lambda}{2}F(M_{\pi}), \\
M_{\pi}^2 &=&
m^2+\frac{\lambda}{6}\phi^2+\frac{\lambda}{6}F(M_{\sigma})
+\frac{5 \lambda}{6}F(M_{\pi}).
\end{eqnarray}
\end{subequations}
Here we have used a shorthand notation and introduced the function
\begin{equation}\label{Eq:FM}
F(M)=\int_{\beta}\frac{1}{\omega_{n}^2+\vec{k}^2+M^2}.
\end{equation}

In order to recover the Goldstone theorem in the truncated CJT effective action, in analogy to the standard One-Particle-Irreducible effective action, a symmetry improved CJT formalism has been proposed in Ref.\cite{Pilaftsis:2013xna} through the introduction of a constraint, which can be consistently implemented by redefining the truncated CJT effective action with a Lagrange multiplier field.  It has the form
\begin{eqnarray}\label{phi2}
\phi M_{\pi}^2=0.
\end{eqnarray}
Substituting this constraint to the above gap equations (\ref{e-mass}), in the symmetry breaking phase, the constrain (\ref{phi2}) implies the massless pions are naturally realized whenever the order parameter is not vanished. On the contrary, in the symmetric phase, we have $\phi=0$ and the constraint is automatically satisfied. As a consequence, the two mass-gap equations become degenerate, the particles have the same mass with $M_{\sigma}=M_{\pi}\equiv M$, and
\begin{equation}\label{Eq:Gap}
M^2=m^2+\lambda F(M).
\end{equation}
In order to get more insight into the nature of the phase transition and verify that the transition is of the second order, We solve the system of mass-gap Eqs.(\ref{e-mass}) with the constraint (\ref{phi2}) using a numerical method based on the Newton-Raphson method of solving nonlinear equations. In this way, we are able to determine the effective masses $M_{\sigma}$ and $M_{\pi}$ and the order parameter $\bar{\phi}$ as functions of temperature $T$.

Using the definition in the framework of the symmetry improved CJT formalism, we can rewrite the finite temperature effective potential $V_{eff}$ as a function of $\psi$ in the thermal Hartree approximation by extending the mass-gap equations (\ref{e-mass}) from $\phi \mapsto \psi$ while
\begin{equation}\label{potential}
M_{\pi}^2 (\psi)=\frac{1}{\psi} \frac{d V_{eff}(\psi)}{d \psi},
\end{equation}
with
\begin{equation}\label{pionms1}
M_{\pi}^2 (\psi)=m^2+\frac{\lambda}{6}F(M_{\sigma})
+\frac{5 \lambda}{6}F(M_{\pi})+\frac{\lambda}{6}\psi^2.
\end{equation}
Furthermore, in chiral limit, in the symmetry breaking phase, $M_{\pi}^2=0$, then
\begin{equation}\label{pionms2}
M_{\pi}^2 (\psi)=-\frac{\lambda}{6}\bar{\phi}^2+\frac{\lambda}{6}\psi^2.
\end{equation}
The order parameter $\bar{\phi}$ as functions of temperature $T$ can analytically expressed in terms of the $T_{c}$ as pointed out in Ref.\cite{Mao:2013gva} as
\begin{equation}\label{sigmamass0}
\bar{\phi}^2=\frac{1}{2}\left( T_{c}^2-T^2 \right)=f_{\pi}^2(1-\frac{T^2}{T^2_c}),
\end{equation}
and the critical temperature $T_{c}=\sqrt{2} f_{\pi}\simeq 131.5  \mathrm{MeV}$. Moreover, the effective mass of sigma meson in the symmetry breaking phase can also described as a function of the  $T_{c}$
\begin{equation}\label{sigmamass}
M_{\sigma}^2=\frac{1}{3}\lambda  \bar{\phi}^2=m^2_{\sigma}(1-\frac{T^2}{T^2_c}).
\end{equation}
Now the Goldstone boson masslessness condition can be naturally implemented in the symmetry breaking phase, and the integration of the effective potential $V_{eff}$ with respect to the field $\psi$ really create a typical potential possessing the second order phase transition.

Integrating the equation (\ref{potential}), we can determine the symmetry-improved effective potential $V_{eff}$ as a function of $\psi$ in Hartree approximation as the form of
\begin{equation}\label{potential2}
 V_{eff}(\psi)=\frac{\lambda}{24} ( \psi^2-\bar{\phi}^2)^2,
\end{equation}
or more specific  in term of the $\varphi$ field as
\begin{equation}\label{potential3}
 V_{eff}( \varphi)=\chi ( \varphi^2-\frac{\bar{\phi}^2}2)^2.
\end{equation}
If we approximately treat a hot thermal medium as a uniform thermal bath with pion strings embedded in it, a new set of coupled equations of motion for the pion string could be derived by simply replacing the relevant classical potential with the above
appropriately modified thermal effective potential  $V_{eff}$. Accordingly, in term of the $\varphi$ field, the equation of motion of $\varphi$ field in Eq.(\ref{pionsf}) can be described as
\begin{equation}\label{pionsf2}
\nabla ^2\varphi =2\chi (\varphi ^{*}\varphi -\frac{\bar{\phi}^2} 2)\varphi .
\end{equation}
From equations (\ref{potential3}) and (\ref{pionsf2}), we can naturally recover the classical potential (\ref{eq2}) and the equation of motion (\ref{pionsf}) for the pion string at zero temperature.

As mentioned in our previous study\cite{Mao:2013gva}, the constraint proposed in Eq.(\ref{phi2}) could be obtained without any assumption in the large-$N$ approximation in the framework of the usual CJT formalism, thus the main results presented above can also be applied to the case of the large-$N$ approximation directly only that the critical temperature $T_{c}=\sqrt{2}f_{\pi}$ is replaced by $T'_{c}=\sqrt{3}f_{\pi}\simeq 161 \mathrm{MeV}$\cite{Petropoulos:1998gt}\cite{Lenaghan:1999si}, and this critical temperature $T'_c$ is closer to the Lattice QCD prediction\cite{Endrodi:2011gv,Bellwied:2015rza,Yagi:2005yb,Fukushima:2010bq}. However, in this work, without any loss of generality, we prefer to treat the critical temperature $T_{c}$ as an input parameter in order to give out a more realistic analysis of the production and evolution of the pion string in high energy heavy ion collisions, e.g. for LHC energies. The reason for this modification is to be as close to the experimental situation as possible.

\begin{figure}[thbp]
\epsfxsize=9.0 cm \epsfysize=6.5cm
\epsfbox{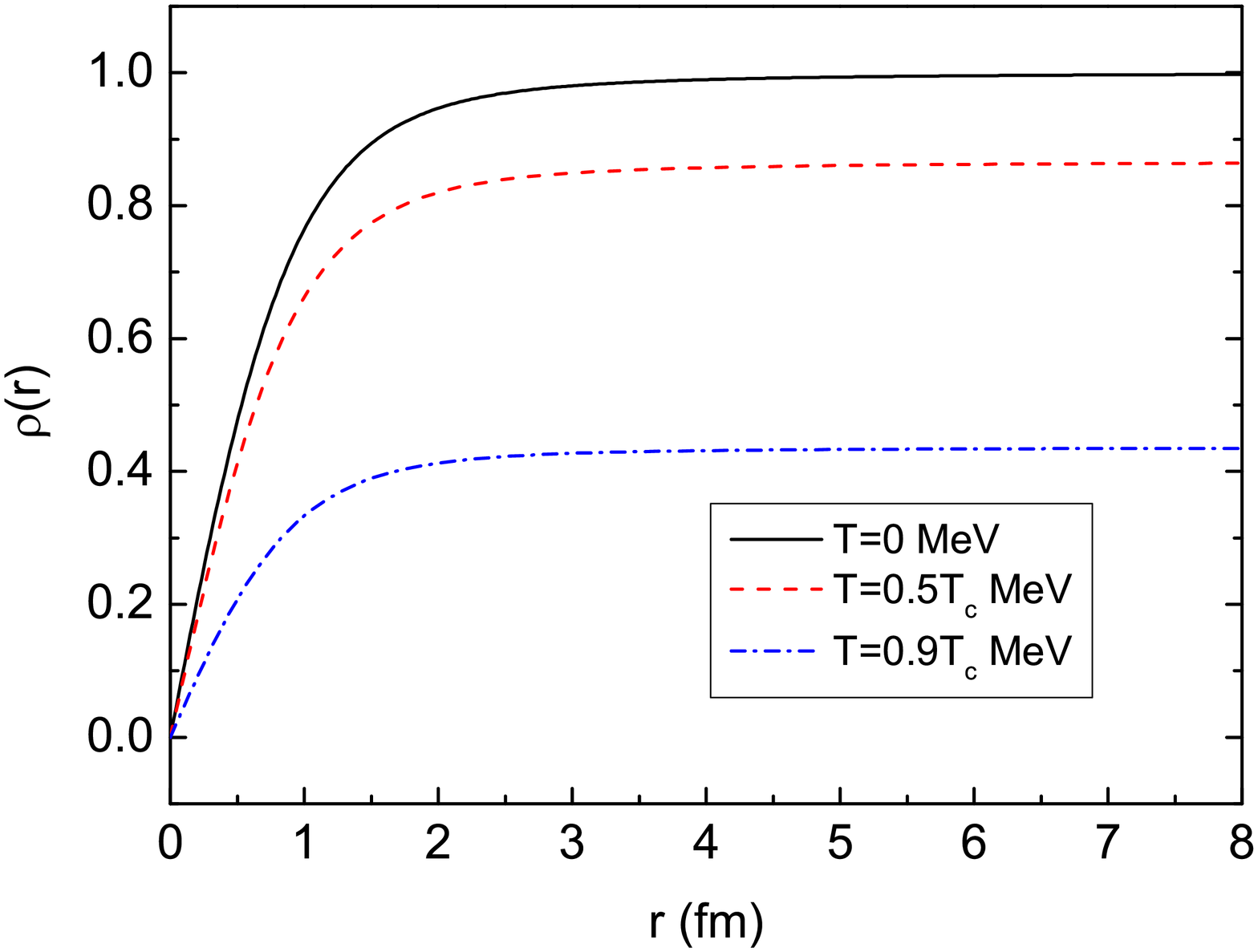}\hspace*{0.1cm} \epsfxsize=9.0 cm
\epsfysize=6.5cm \epsfbox{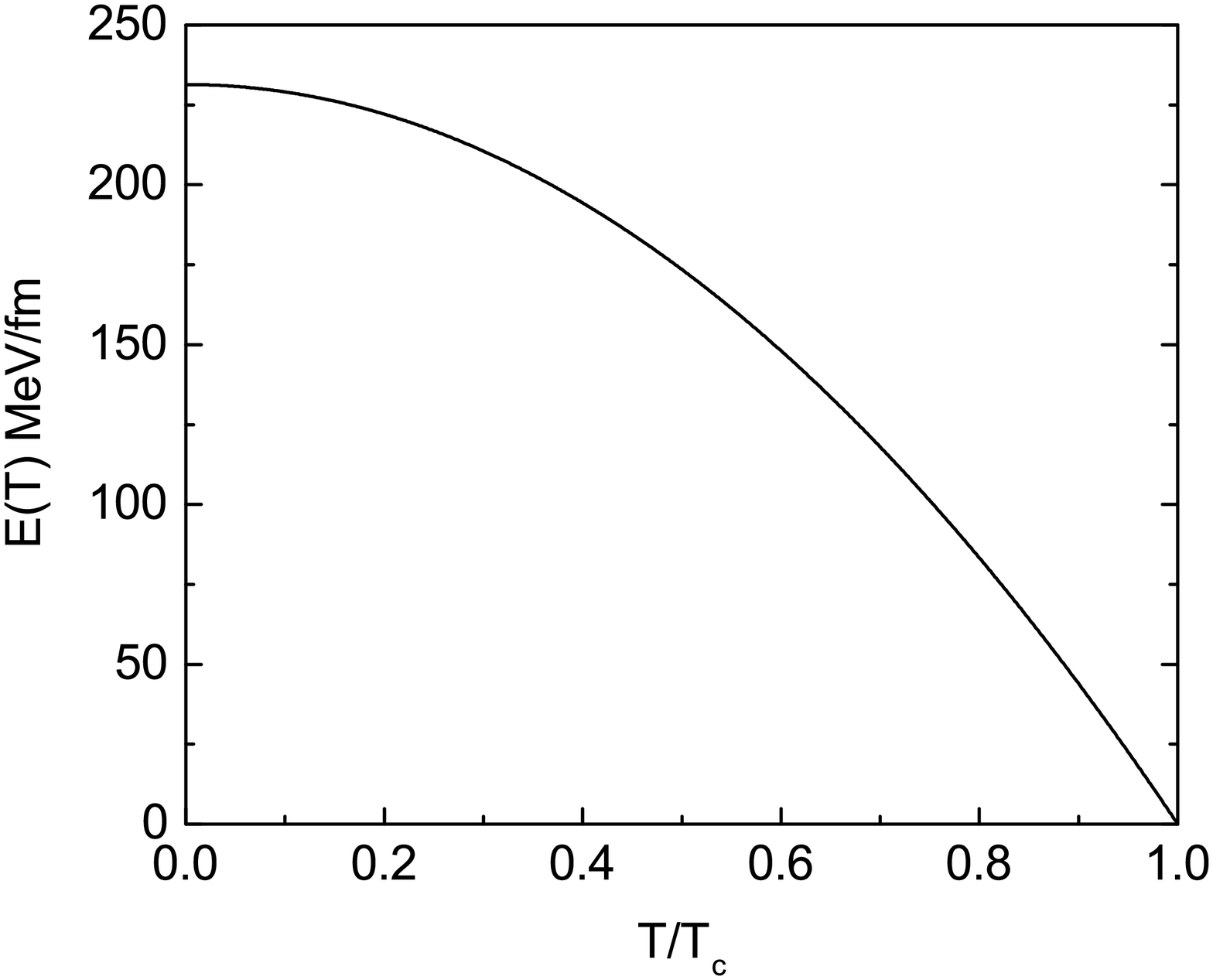}
\vskip -0.05cm \hskip 0.15 cm \textbf{( a ) } \hskip 6.5 cm \textbf{( b )} \\
 \caption{Numerical solution of the pion string in the case of chiral limit with $n=1$. (a) The profile functions of pion strings as function of the radius $r$ at various temperatures, where the solid curve is for zero temperature, the dashed curve is for $T = 0.5  T_{c}$  and the dash-dotted curve is for $T = 0.9 T_{c}$ . (b) The energy densities as a function of the temperature $T$. }
\label{Fig01-02}
\end{figure}

We first numerically investigate pion string solutions at finite temperature. In Fig.\ref{Fig01-02}, we plot the profile functions of the pion string for different temperatures. It is shown that all the fields are moving towards to the trivial values with increasing temperature. When $T$ is larger than some critical temperature $T_{c}$, there exist only the trivial solutions for the equation of motion in Eq.(\ref{pionsf2}), and the pion string is melted away. Correspondingly, the trivial solution indicates the restoration of chiral symmetry in full space.

\section{The Kibble-Zurek mechanism}
In order to study whether the pion string exists during the chiral phase transition, we have to estimate how many pion strings are created in the phase transition by the Kibble-Zurek mechanism\cite{Kibble76,Zurek:1996sj,Mao:2004ym,delCampo:2013nla}. As described by the Kibble-Zurek mechanism, the pion strings can be expected to form in a second order phase transition even it is a perfectly homogeneous fast transition, if the transition proceeds faster than the order parameter of the breaking symmetry phase is able to relax. In such a non-equilibrium transition, because of fluctuations of the order parameter, the new symmetry-broken-phase starts to form simultaneously and independently in many parts of the system. Subsequently, these independent regions grow together to form the new symmetry-broken phase randomly. At the boundaries, where different causally disconnected regions meet, the order parameter does not necessarily match and a domain structure is formed. Such a random domain structure reduces to a network of strings if the broken symmetry is the U(1)  symmetry.

Facilitated with the Kibble-Zurek mechanism, we consider a system cooling from an initial temperature $T_i$ through the critical temperature $T_{c}$ by the change of a control parameter $\varepsilon$. For quantitative estimation, we take a quark-gluon plasma created at the LHC Pb-Pb collisions with a beam energy at $\sqrt{s_{NN}}=2760 \mathrm{GeV}$ as a thermal bath. As a result of ultrarelativistic collision between two heavy ions, a fireball is expected to form and then it will rapidly thermalize due to the fact that there exists manifestly partons with very large cross section for gluon scattering. The enormous amount of energy density deposited in the fireball results in large pressure gradients from the central to the peripheral region of the fireball that drives the expansion of the fireball. This expansion leads to cooling of the fireball and the hydrodynamic model gives that the temperature of the fireball $T$ evolves with the time $t$ as\cite{Bjorken:1982qr,Schnedermann:1992hp,Hung:1997du,Heinz:2000ba,Andronic:2011yq}
\begin{equation}\label{tempvt}
T(t) \propto {t}^{\frac13}.
\end{equation}

Furthermore, with the expansion of the fireball, the interparticle distance grows with time (or temperature), the particles cease to interact after sometime and free stream to the detector. The surface of last scattering is called as the freeze-out surface. Since scattering could be both elastic (where particle identities do not change) and inelastic (where particle identities change), it is possible to have two distinct freeze-outs, namely, chemical freeze-out, where inelastic collisions cease, and thermal/kinetic freeze-out where elastic collisions cease and the particle mean free path becomes higher than the system size, which forbids the elastic collision of the constituents in the system\cite{Heinz:2004qz}. In experiment, the chemical freeze-out surface is determined by analysing the measured hadron yields, while the kinetic freeze-out surface can be determined by studying the data of transverse momentum distribution of produced particles. In Ref.\cite{Chatterjee:2015fua}, the chemical and kinetic freeze-out scenarios in central heavy ion collisions from the lower AGS energies to the largest LHC energies are investigated in detail. Based on their studies, the extracted chemical and kinetic freeze-out temperature corresponding to LHC energies for zero chemical potential are $T_{{ch}}\simeq150 \mathrm{MeV}$ and $T_{{kin}}\simeq 90 \mathrm{MeV}$, respectively. The radius which is reflective of volume of the fireball can also extracted as $r_{{ch}}\sim 10.87 \mathrm{fm}$ at $T_{\mathrm{{ch}}}\simeq150 \mathrm{MeV}$. Correspondingly, the volume of the fireball at the moment is about $V(T_{{ch}})\simeq 5380 \mathrm{fm}^3$ and the subsequent evolution of the volume is given by
\begin{equation}
V(t) \propto {t}.
\end{equation}

Employed with the resulted thermal parameters, temperature $T_{{ch}}$, we can now fix the critical temperature $T_{c}$. It was realized that there are very high densities in the system when the temperature is close to the critical temperature $T_{c}$, but as long as the temperature is below the critical temperature, the quark-gluon plasma created at heavy ion collisions experiences a hadronization, and the
rapid fall-off of density lead to rapid equilibration and chemical freeze-out only a few MeV below the transition temperature\cite{BraunMunzinger:2003zz}.  The current understanding of the connection between the QCD phase diagram and chemical freeze-out can be revealed through the comparison of lattice calculations with results from the hadronic gas models. Therefore, from the recent work of \cite{Bellwied:2015rza}, the precise critical temperature is $T_{c}\simeq 157\mathrm{MeV}$ for zero chemical potential.

In the vicinity of  $T_{c}$, a continuous second-order phase transition can be characterized by the divergence of both the equilibrium correlation length $\xi$
\begin{equation}
\xi(\varepsilon)=\frac{\xi_0}{\varepsilon^{\nu}},
\end{equation}
and equilibrium relaxation time $\tau$
\begin{equation}
\tau(\varepsilon)=\frac{\tau_0}{\varepsilon^{z\nu}},
\end{equation}
as a function of the distance to the critical point $T_{c}$. Where $\xi_0$ is the zero temperature-limiting value of the temperature-dependent coherence length $\xi(T)$, and in this work we take $\xi_0=1/m_{\sigma}\simeq 0.4 \mathrm{fm}$. But for $\tau_0$, since the experimental data obtained in the RHIC and LHC experiments favor a very short thermalization/equilibration time, we take $\tau_0 \sim 0.06  \mathrm{fm}$ so as to gain a reasonable value for the critical time $t_c$ at $T_{c}\simeq 157\mathrm{MeV}$ for zero chemical potential. From Eq.(\ref{sigmamass}), the correlation length $\xi(T)$ is given by $\xi(T)=(\frac{1}{3}\lambda \bar{\phi}^2)^{-1/2}$, where $\bar{\phi}$ is the global minimum of the effective potential in the homogeneous case \cite{Yagi:2005yb}. Consequently, it is convenient to define our reduced distance parameter as
\begin{equation}
\varepsilon=\frac{T^2_c-T^2}{T^2_c}=1-\frac{T^2}{T_{c}^2},
\end{equation}
and we can get the correlation length critical exponent $\nu=1/2$. While the dynamic critical exponent $z$ is  equal to $2$ in a Ginzburg-Landau system at a second order phase transition. In terms of the reduced distance parameter $\varepsilon$, the system initially prepared in the high-symmetry phase ($\varepsilon<0$) is forced to face a spontaneous symmetry breaking scenario as the critical point is crossed towards the degenerate vacuum manifold ($\varepsilon>0$).

The reduced parameter is characterized by the quench time $\tau_Q$,
\begin{equation}
\tau_Q=\left( \frac{1}{T_{c}} \left| \frac{dT}{dt}  \right|_{T=T_{c}} \right)^{-1}=3t_{c},
\end{equation}
and varies linearly in time $\varsigma$ according to
\begin{equation}\label{tauq}
\varepsilon=\frac{\varsigma}{\tau_Q},
\end{equation}
in $\varsigma\in \left[-\tau_Q,\tau_Q \right]$,  the critical point being reached at $\varsigma = 0$ and the system time $t$ is given by $t=t_{c}+\varsigma$. Far away from the critical point $T\gg T_{c}$ or $T\ll T_{c}$, the equilibrium relaxation time is very small with respect to the time remaining until reaching the critical point following the quench (\ref{tauq}), and the dynamics is essentially adiabatic. In the opposite limit, in the close neighborhood of $\varepsilon= 0$, the dynamics is approximately frozen due to the divergence of the equilibrium relaxation time (critical slowing down). The system is then unable to adjust to the externally imposed change of the reduced control parameter $\varepsilon$, and the order parameter of the system ceases to follow the equilibrium expectation value and enters an impulse stage within the time interval $[-t_{z},t_{z}]$. This is often referred to as the adiabatic-impulse approximation, which is believed to capture the essence of the nonequilibrium dynamics involved in the crossing of the phase transition at a finite rate. The inability of the order parameter to keep up with the change imposed from the outside is the essence of the freeze-out.

The above time boundary between the adiabatic and frozen stages $t_{z}$ can be estimated as follows. When $\varsigma>0$ or the temperature is below $T_{c}$, the order parameter coherence spreads out with the velocity
\begin{equation}
c(T) \sim \frac{\xi}{\tau}=\frac{\xi_0}{\tau_0}\varepsilon^{\frac12}.
\end{equation}
The freeze-out of pion strings are expected to occur at the Zurek time $t_{z}$ when the causally disconnected regions have grown together and the system is to establish coherence in the whole volume. In the same time, the global thermodynamic equilibrium of the system may be realized and the order parameter will follow its equilibrium value in whole space. Therefore, this moment is also corresponding to the end of the hadronization process in heavy ion collisions and the freeze-out temperature $T_{z}$ at the Zurek time $t_{z}$ in the Kibble-Zurek mechanism can be directly identified as the chemical freeze-out temperature $T_{\mathrm{{ch}}}$ due to the fact that the very hadronization process leads to the equilibrium distributions of hadrons\cite{Heinz:2000ba}.

\begin{figure}
\includegraphics[scale=0.36]{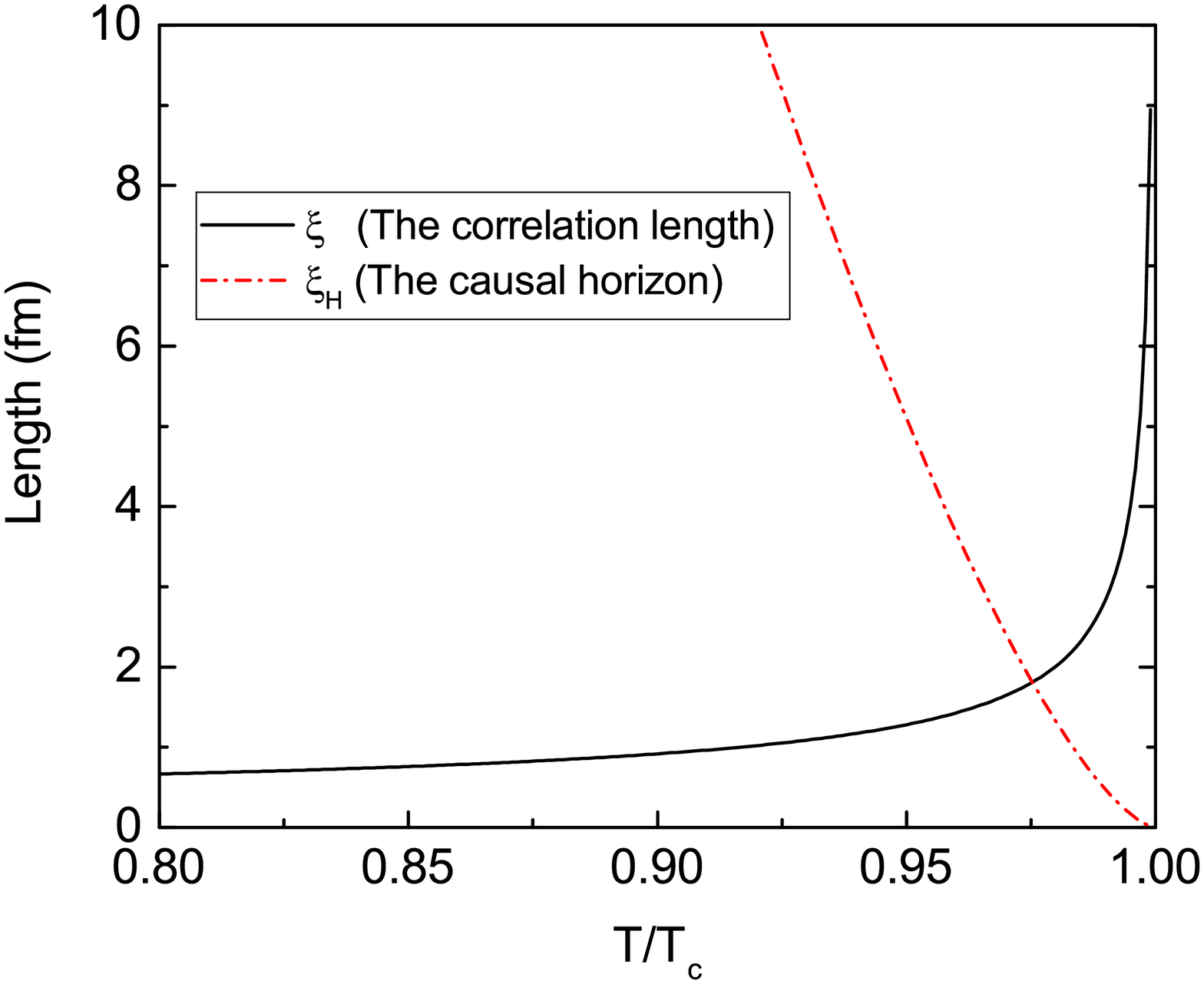}
\caption{\label{Fig03} With decrease of temperature $T\leq T_{c}$, the causal horizon will move out to a distance $\xi_H$, and it will be subsequently equal to the coherence length $\xi (\varsigma_z)$ when the equilibrium phase is achieved at the freeze-out point $T_{z}\simeq 150 \mathrm{MeV}$.}
\end{figure}

 At the freeze-out temperature $T_{z}$ ($T_{z}=T_{\mathrm{{ch}}}$), the causal horizon has increased to the distance
\begin{equation}
\xi_H(\varsigma_z)=\int^{\varsigma_z}_0 c(T)d\varsigma=\frac{\xi_0 \tau_Q}{\tau_0} \varepsilon^{\frac32}_z,
\end{equation}
which has to be equal to the coherence length $\xi(\varsigma_z)$ as shown in Fig.\ref{Fig03}. This condition leads to a relationship between the freeze-out temperature $T_{z}$ and the freeze-out time $\varsigma_z$ as follows:
 \begin{eqnarray}
\varepsilon_z = \sqrt{\frac{\tau_0}{\tau_Q}}= \frac{\varsigma_z}{\tau_Q}.
\end{eqnarray}
Then, it is easy to obtain
\begin{eqnarray}
t_{z}=t_{c}+\varsigma_z=t_{c}+\sqrt{\tau_0 \tau_Q}\simeq 9.60 \mathrm{fm},
\end{eqnarray}
when temperature is at $T_{z}\simeq 150 \mathrm{MeV}$. Here, $t_c=3\tau_0/\left(\left({T_c}/{T_{z}}\right)^3-1 \right)^2\simeq 8.37 \mathrm{fm}$, which is consistent with the calculation of the hydrodynamic models used to interpret the RHIC and LHC results\cite{Florkowski:2010zz}.

The Kibble-Zurek mechanism sets the average size of the domains by the value of the equilibrium correlation length at $\varepsilon_z$,
\begin{equation}
\xi_z=\xi(\varsigma_z)=\xi_0(\frac{\tau_Q}{\tau_0})^{\frac14}\simeq 1.81 \mathrm{fm}.
\end{equation}
Moreover, it is worth to point out that the equilibrium correlation length at $\varepsilon_z$ should be taken as the horizon size or the typical separation length between strings $R$. In general,  the correlation length $\xi_z$ is often recast as an estimate for the resulting density of topological defects,
\begin{equation}\label{density}
\mathfrak{n}=\frac{1}{\kappa \xi_z^2},
\end{equation}
where the parameter $\kappa$ is introduced to reduce the overestimation of real density of defects in experiment and $k \sim 1-100$ depends on the details of the specific model. In this study, the experimental measurements in heavy-ion collisions will give a constraint on $\kappa$. $\mathfrak{n}$ is defined as string length per unit volume. We can estimate $\mathfrak{n} \simeq 0.3 \mathrm{fm}/\mathrm{fm}^3$ for $\kappa=1$ and  $\mathfrak{n} \simeq 0.003 \mathrm{fm}/\mathrm{fm}^3$ for $\kappa=100$. This is the main prediction of  the Kibble-Zurek mechanism, and the result presented here can be generalized to other Ginzburg-Landau systems having a second-order phase transition.

\section{The decay of the pion string}

Note that in our situation, closed string loops give dominant contributions to the total energy of the strings. This is because at the freezing out time $t_{z}$, the string evolution is still ruled out by the frictional force by the surrounding matter so that the free motion of the string is not realized. In addition, the expansion of the system is too rapid that the initial Brownian string distribution will be conserved. Then the initial structure of the string network partially remain and the spatial trajectories of strings are very complicated. Thus we take the initial pion string network as that of the Brownian one and regard that the distribution of these loops does not change with time $t\leq t_z$ so long as they could survive in such a ruinous  environment, the size of the loops are conformally stretched during the expansion of fireball and a simple scaling can be realized.

By using the scale invariance of Brownian string described by Vachaspati and Vilenkin\cite{Vachaspati:1984dz}, the initial distribution of number density of the pion string with the length between $l$ and $l + dl$ at the chemical freeze-out temperature $T_{{ch}}$ can be described as
\begin{eqnarray}\label{stringd}
dn(l)=k\xi_z^{-\frac32}l^{-\frac52}dl,
\end{eqnarray}
where the parameter $k$ is not a free parameter and it is easy to establish the relationship between the above parameter $\kappa$ in equation (\ref{density}) with the parameter $k$ here. As we already know $\mathfrak{n}$ is the string length per unit volume and the volume of the fireball at $T_{z}$ is $V(T_z)\simeq 5380 \mathrm{fm}^3$, then the total length of the pion string in the fireball at the freeze-out time $t_z$ is set by $L_T\simeq \mathfrak{n} \times V(T_z)$. On the other hand, as long as we obtain the distribution of number density of the pion string $n(l)$, the total length of the pion string in the fireball at the freeze-out time $t_z$ is  given by
\begin{eqnarray}\label{tlength}
L_T=V(T_z)\int_{l_a}^{l_b} l dn(l)=kV(T_z)\xi_z^{-\frac32} \int_{l_a}^{l_b} l^{-\frac32}dl=2kV(T_z)\xi_z^{-\frac32}(l_a^{-\frac12}-l_b^{-\frac12}).
\end{eqnarray}
Here, $l_a$ and $l_b$ are the minimum and maximum lengths of the pion strings, respectively. Note that the string width $\xi_z$ gives a minimum length of the pion string at time $t_z$, $l_a\simeq 2\pi\xi_z\simeq 11.37\mathrm{fm}$, while the longest string is also constrained by the volume of the system with $l_b\simeq 2\pi r_{ch} \simeq 68.26\mathrm{fm}$. After a simple calculation, we arrive at
\begin{eqnarray}
k=\frac{1}{2\kappa \xi_z^{\frac12}(l_a^{-\frac12}-l_b^{-\frac12})}.
\end{eqnarray}
Then if $\kappa=1$, $k$ is about $2.12$, whereas $k$ is about $0.02$ for $\kappa=100$.

After integration of the equation (\ref{stringd}), we have the total number of pion strings:
\begin{eqnarray}
N(T_z)=V(T_z)\int_{l_a}^{l_b} dn(l)=\frac{2kV(T_z) }{3 \xi_z^{\frac32}} \left(l_a^{-\frac32}-l_b^{-\frac32}\right)\simeq 58k.
\end{eqnarray}
Then the total number of pion strings created at the LHC Pb-Pb collisions with a beam energy at $\sqrt{s_{NN}}=2760 \mathrm{GeV}$ could vary between $N\simeq 116$ and $N\simeq 1$.

Unfortunately, in order to detect these produced pion strings in the heavy ion collisions experiments, we need to deliver a comprehensive analysis on the process of the pion string decay after the time $t_z$.  For simplicity and convenience, we separate our subsequent studies into two cases which are corresponding to two limits of the stabilities of pion strings. The first case is that the produced pion strings are suggested to be absolutely stable during the time from the chemical freeze-out  time $t_z$ to the kinetic/thermal freeze-out time $t_f$ according to the plasma stabilization mechanism proposed in Refs.\cite{Nagasawa:1999iv,Nagasawa:2002at,Karouby:2012yz,Karouby:2013vza}. But, as soon as the time is across $t_f$, the particles in hadron resonance gas depart from each other so fast that the collision processes become ineffective, thus pion strings become unstable and decay suddenly. On the contrary, there exists another possibility: the plasma stabilization mechanism does not work very well in the case of the heavy ion collisions experiment, and the produced pion strings during the chiral phase transition will decay immediately after their births when $t>t_z$.  We will point out later that the second case is closer to the experimental situation.

\subsection{The stable case}
In this case, after the freeze-out temperature $T_{z}$, the pion string will stop forming and scale with the expansion of the system or the fire ball created in the heavy ion collisions until to the kinetic freeze-out time $t_f$. Because the volume of the system obeys the following law $V(t)\sim t$, then we make an assumption that the correlation length of the pion string will rescale as
\begin{equation}\label{scaling}
\xi_{t}=\xi_z(\frac{t}{T_{z}})^{\frac13},
\end{equation}
until to the finial stage of the evolution of the fire ball. Based on the chemical and kinetic freeze-out scenarios in Ref.\cite{Chatterjee:2015fua}, the kinetic freeze-out temperature will occurs at $T_f\simeq 90 \mathrm{MeV}$, so that we have $t_f\simeq 44.43 \mathrm{fm}$. Now the correlation length of the pion string achieves at $\xi_{f}\simeq 3.02 \mathrm{fm}$, which gives out the minimum length of the pion string at time $t_f$
as $L_{f}=2 \pi \xi_{f}\simeq 18.97 \mathrm{fm}$.

By making the assumption that pion strings are absolutely stable in a plasma and can survive after the decoupling time $t_f$, the evolution and decay of the formed pion string has been extensively investigated in Ref\cite{Mao:2004ym}. The qualitative results obtained in that work are applied in this work too in the case of slight modification of the parameters over there. In order to avoid the simple repetition, in this section, we only bring forward some comments on the previous work presented in Ref\cite{Mao:2004ym}.

The first comment: the thermal or kinetic freeze-out is the stage in the evolution of matter when the hadrons practically stop to interact, at that time, the transverse momentum distributions of the particles get fixed and do not change thereafter. If pion strings can survive after the decoupling time $t_f$, neutral pions and sigma (charged pions) particles emitted from pion strings do not get chance to be thermalized. These unthermalized pions will give a reasonable contribution to the $p_T$ spectra of $\pi$ and induce a visible fluctuation/distortion in the low $p_T$-momentum regimes. In fact, the experimental data show the hadron spectra in the soft region, $p_T\leq2 \mathrm{GeV}$, have thermal character, although the original thermal distributions are modified by the collective transverse flow and decays of resonances. So that if pion strings are really formed during the QCD chiral phase transition, they should decay completely before $t_f$. Otherwise, they could be already observed in experiment.

The second comment: based on the chemical and kinetic freeze-out scenarios in Ref.\cite{Chatterjee:2015fua}, the chemical freeze-out time $t_z$ is the moment when the inelastic collisions stop, so that particle identities do not change and the temperature of the freeze-out is inferred from the studies of the ratios of hadron multiplicities. Moreover, when $t\geq t_z$, the hadron resonance gas is taken as a gas of noninteracting hadrons and resonances which is a good low $T_{ch}$ approximation to QCD thermodynamics. Any additional pions burst from pion strings when $T<T_{ch}$ would pose a serious change to the standard chemical freeze-out models which have been quite successful in describing the particle multiplicities measured in heavy-ion collision experiments. Therefore, it is more reasonable to believe that pion strings will decay immediately soon after they are produced in heavy ion collisions.

The third comment: the plasma stabilization mechanism proposed in Refs.\cite{Nagasawa:1999iv,Nagasawa:2002at,Karouby:2012yz,Karouby:2013vza} does work only when the coupling constant $\lambda$ is extremely small. For a realistic value of $\lambda$ in this work, the temperature condition for the stability of the pion string is
\begin{equation}
T>T_D=\frac{2\sqrt{\chi}}{e}f_{\pi}\simeq 2335 \mathrm{MeV}.
\end{equation}
$T_D$ is even much larger than the $T_c$, so it is natural that pion strings produced in LHC experiment are unstable and they will decay quickly after giving birth.

\subsection{The unstable case}
In the Kibble-Zurek mechanism, the fast second-order chiral phase transition of the fireball leads to the formation of pion strings at freeze-out time $t_z$. In analogy to a cosmic string created in cosmology\cite{Villenkin00}, the pion strings formed in nucleon-nucleon collisions may be imagined as sigma - neutral pion pairs confined in a string-like tube. In the next step, the strings decay/fragment forming directly hadrons (the sigma particles and neutral pions), and the produced hadrons are modeled as smaller pieces of the original string.

As mentioned above, the pion string is not topologically stable, since any field configuration can be continuously deformed to the vacuum. It can be shown by numerical analysis that the pion string is only stable for very small values of $\lambda$. This implies that the pion string is unstable in the parameter space allowed experimentally\cite{Zhang:1997is}. In the early universe and in heavy-ion collisions, pion strings are expected to be produced and subsequently decay. Their lifetime can be estimated by considering their interactions with the surrounding plasma. Based on a naive dimensional analysis, their lifetime $\tau_{ps}$ should be proportional to the inverse of the their masses/energies at specific temperature.

From the right panel in Fig.\ref{Fig01-02}, the energy per unit length of the pion string can be parameterized as
\begin{equation}
E(T)=E_0 (1-\frac{T^2}{T_{c}^2}),
\end{equation}
with $E_0$ is the energy of the pion string at vacuum, and $E_0\simeq 231.38 \mathrm{MeV/fm}$. At $T=T_z$, the energy per unit length of the pion string is about $20.17 \mathrm{MeV/fm}$. Since a minimum length of the pion string at time $t_z$ is $l_a\simeq 2\pi\xi_z\simeq 11.37\mathrm{fm}$, then the mass of the pion string $M(T)$ at $T=T_z$ at least has
\begin{equation}
M(T_z)=2\pi\xi_z E(T) \simeq 229.3 \mathrm{MeV}.
\end{equation}
Correspondingly, the lifetime $\tau_{ps}$ of the pion string at moment is less than $0.86 \mathrm{fm}$. Thus we can see the pion string indeed decay immediately soon after they are produced in heavy ion collisions at $t=t_z$.

At chemical freeze-out, the inelastic interactions among the produced particle stop. However, these particles can still interact elastically which could affect their momentum distributions. Then all sigmas and pions burst from pion strings will be thermalized quickly in a plasma. Furthermore, since the thermal character in the hadron spectra in experimental data can not be modified by the collective transverse flow and decays of resonances. It is extremely difficult to recognize and isolate sigmas and pions decayed by pion strings from the background thermal productions by using the efficient method of the transverse momentum distributions or the HBT interferometry (more recently, the name femtoscopy is used) technique\cite{Wiedemann:1999qn}.

Even though the produced pion string is unstable, there still have possible signals of the pion string produced because the string configuration violates the isospin symmetry. Since the generic expectation is that pions of different isospins would be produced in equal abundances because of isospin symmetry, however, the formation and decay of pion strings would produce abundant neutral pions, which would lead to a large deviation in the charge-neutral correlation from expectations based on the generic pion production mechanism. Similarly to the DCC (disoriented chiral condensate) phenomenon\cite{Blaizot:1992at,Rajagopal:1992qz,Mohanty:2005mv}, pion strings are also expected to emit pions coherently from the collision volume, which may results in large fluctuations (sometime called as DCC-like fluctuations) in the neutral pion fraction, $f$, defined as $f= n_0/(n_0 + n_c)$, where $n_0$ and $n_c$ are the multiplicities of the neutral and charged pions, respectively.

Experimentally, the existence of DCC phenomenon was previously investigated in heavy-ion collisions at the CERN Super Proton Synchrotron at $\sqrt{s_{NN}}=17.3 \mathrm{GeV}$ \cite{Aggarwal:1997hd,Aggarwal:2000aw,Aggarwal:2002tf,Collaboration:2011rsa} and at Tevatron in $p+\bar{p}$ collisions by the MiniMax collaboration at $\sqrt{s_{NN}}=1.8 \mathrm{TeV}$\cite{Brooks:1999xy}\cite{Brooks:1996nu}.In both cases, the possibility of large-sized DCC domain formation has been excluded by the measurements. Recently, event-by event fluctuations of the multiplicities of inclusive charged particles and photons at forward rapidity in $Au+Au$ collisions at $\sqrt{s_{NN}}=200 \mathrm{GeV}$ have been studied in Ref.\cite{Adamczyk:2014epa}, in their studies, it is shown that for all pions, the observational deviation is found to be less than $1\%$ from the generic expectations based on the generic production mechanism of pions owing to isospin symmetry.This means that if pion strings or DCC domain are really produced in the heavy-ion collisions, the number of the produced pions should be less than $1\%$ of the total pions in fireball. Under this constrain, at the LHC Pb-Pb collisions with a beam energy at $\sqrt{s_{NN}}=2.76 \mathrm{TeV}$\cite{Abelev:2013vea}, the numbers of sigmas and neutral pions produced from pion strings are expected to be $N_{\pi^0}\sim N_{\sigma}\sim 25-7$ if we ignore the effect of the DCC phenomenon or other effects. Then the total number of pion strings created at the LHC Pb-Pb collisions are about $N\simeq 4$ and $N\simeq 1$, which give $k \in [0.02,0.07]$. Therefore, the production rate of topological defects seems much lower than that of system in condensed matter physics\cite{Eltsov:1998fv}.

\section{Summary and discussion}
We have investigated the production and evolution of the pion strings in a thermal bath within the framework of the $O(4)$ linear sigma model in chiral limit by adopting the symmetry-improved CJT effective formalism. It has been shown that the formalism is keeping the Goldstone theorem and satisfying a second-order phase transition, this makes it possible to apply the Kibble-Zurek mechanism to study the formation and evolution of the pion string in a chiral phase transition in early universe or heavy ion collision. Our results indicate that pion strings are expected to be produced in a second-order chiral phase transition, and then quickly decay into neutral pions and sigmas after the chemical freeze out time $t_z\simeq 9.6 \mathrm{fm}$. Following the Kibble-Zurek mechanism, we can roughly estimate the number of pion strings are to be $1-4$, and their subsequent decays will produce about $7-25$ neutral pions and sigmas. Because the string configuration violates the isospin symmetry, the neutral pions are more likely to be produced during the chiral phase transition when comparing to the charged pions. This can be taken as a possible signal of the pion string.

However,  the possible signal observed in this work is very similar to the experimentally observable anomalies caused by the DCC decay. A further effort needs to be addressed in future in order to clarify the differences from the pion string decay and the DCC decay. Another problem deserved to be noted is that chiral phase transition is actually a crossover rather than a phase transition, then the estimation of the pion string number density is not so straightforward by directly using the Kibble-Zurek mechanism, and simulation of the production and evolution of pion strings in early universe or heavy ion collision remains as an important future problem. These two problems are certainly out of the scope of our current topic and we prefer leaving for future study.

Before concluding, let us give some notes about the existence of topologically stable non-Abelian global strings in the $U(N)$ linear sigma model\cite{Nitta:2007dp,Eto:2009wu,Eto:2013hoa}. If the axial anomaly is absent, there exist stable Abelian axial string (or $\eta$ string\cite{Zhang:1997is}) winding around the spontaneously broken $U(1)_A$ and non-Abelian axial string winding around both the $U(1)_A$ and non-Abelian $SU(N)$ chiral symmetries. Therefore, If the $U(1)_A$ symmetry is exact above the QCD chiral
phase transition, then, as long as the effects of instantons can be neglected, the $\eta$ string and non-Abelian axial string can form during the chiral phase transition of QCD, and moreover, they are topologically stable. This is unlike the pion string, these produced Abelian and non-Abelian axial strings can survive until to the kinetic freeze-out time $t_f$, when $t>t_f$, the particles depart from each other very fast and the collision processes become ineffective. Thereafter, large pressure gradients from the central to the peripheral region of fireball suddenly switch off and these stable strings start to shrink to their minimum lengths then decay quickly. Moreover, as the density of matter and/or the temperature decreases, it is expected that the instanton effects will rapidly appear and the axial anomaly term has to be included in the model. Therefore, metastable domain walls are present and Abelian axial strings must be attached by $N$ domain walls, forming domain wall junctions\cite{Eto:2013bxa}\cite{Li:2004yw}. Thenceforth, Instable domain wall junctions will decay sooner or later through quantum and thermal effects. It is interesting to investigate whether domain walls and strings are stable or not until the kinetic freeze-out time $t_f$ since they seem to live much longer than pion strings. And we believe any kind of topological solitons created in the heavy ion collisions will be interested both for the theoretical and experimental studies.  Because these formed stable solitons can be directly detected and confirmed by the experimental observation if they can live on until the kinetic freeze-out time $t_f$. Eventually, work in this direction is under progress.

\begin{acknowledgments}
We thank Jingjing Bao, Jinshuang Jin, Daicui Zhou and Rongxiang Zhu for valuable comments and discussions. The project is Supported in part by NSFC under No.11274085 and 11275002.
\end{acknowledgments}

\end{document}